\documentclass[aps,pra,reprint,amsmath,amssymb,superscriptaddress,onecolumn,longbibliography,nofootinbib,notitlepage]{revtex4-2}
\usepackage{mathtools}
\usepackage{siunitx}
\usepackage[hidelinks]{hyperref}
\usepackage{svg}
\svgpath{{Figures/}}
\usepackage{caption}
\usepackage{xfrac}
\usepackage{float}
\usepackage{subcaption}
\usepackage[braket, qm]{qcircuit}
\usepackage{bbm}
\usepackage{lipsum} 

\usepackage{soul}
\usepackage[normalem]{ulem}

\usepackage{xcolor}

\newcommand{\correspondingauthors}{%
To whom correspondence should be addressed:  ps2228@cornell.edu, mo522@cornell.edu, and pmcmahon@cornell.edu.}

\usepackage{ragged2e}

\makeatletter
\long\def\@makecaption#1#2{%
  \par
  \vskip\abovecaptionskip
  \small
  \justifying
  \noindent #1. #2\par
  \vskip\belowcaptionskip
}
\makeatother

\begin{document}
\title{Experimental quantum-computing-enhanced sensing using Grover's algorithm}

\author{Mathieu~Ouellet}
\thanks{These authors contributed equally.}
\affiliation{School of Applied and Engineering Physics, Cornell University, Ithaca, NY 14853, USA}

\author{Purnendu~Sen}
\thanks{These authors contributed equally.}
\affiliation{School of Applied and Engineering Physics, Cornell University, Ithaca, NY 14853, USA}

\author{Xiangqin~Wang}
\affiliation{School of Applied and Engineering Physics, Cornell University, Ithaca, NY 14853, USA}

\author{Saswata~Roy}
\affiliation{School of Applied and Engineering Physics, Cornell University, Ithaca, NY 14853, USA}
\affiliation{Department of Physics, Cornell University, Ithaca, NY 14853, USA}

\author{Xingrui~Song}
\affiliation{School of Applied and Engineering Physics, Cornell University, Ithaca, NY 14853, USA}

\author{Vladimir~Kremenetski}
\affiliation{School of Applied and Engineering Physics, Cornell University, Ithaca, NY 14853, USA}

\author{Sridhar~Prabhu}
\affiliation{School of Applied and Engineering Physics, Cornell University, Ithaca, NY 14853, USA}
\affiliation{Department of Physics, Cornell University, Ithaca, NY 14853, USA}

\author{Valla~Fatemi}
\affiliation{School of Applied and Engineering Physics, Cornell University, Ithaca, NY 14853, USA}

\author{Peter~L.~McMahon}
\thanks{\correspondingauthors}
\affiliation{School of Applied and Engineering Physics, Cornell University, Ithaca, NY 14853, USA}

\begin{abstract}
The combination of quantum sensing with quantum computing to provide an enhancement over conventional quantum sensing has recently emerged as a promising potential application of quantum computing that could give advantages without needing large-scale or fault-tolerant hardware. In this work, we report an experimental demonstration of a recent theoretical proposal to repurpose Grover's search algorithm to improve the ability to detect signals with unknown frequency within a large detection bandwidth. Our experiments were based on a system comprising a single superconducting qubit coupled to a single superconducting cavity, highlighting the modest hardware requirements for realizing the protocol. We found that Grover-based sensing was able to outperform the natural non-Grover baseline for our experimental platform for detection bandwidths $>$10~MHz, with an advantage that empirically grew superlinearly with the bandwidth beyond that break-even point. The use of the Grover-based protocol reduced the amount of signal that needed to be sensed to make an accurate detection decision by more than 10$\times$ for choices of larger detection bandwidth and higher desired detection accuracy. Our results provide a proof-of-principle validation that Grover-based quantum computational sensing can be realized in near-term hardware and provide a metrological advantage well beyond break-even in spite of the additional protocol complexity.
\end{abstract}

\maketitle

\section{Introduction}
\label{sec:intro}
\begin{figure*}
    \centering
    \includegraphics[width=\linewidth]{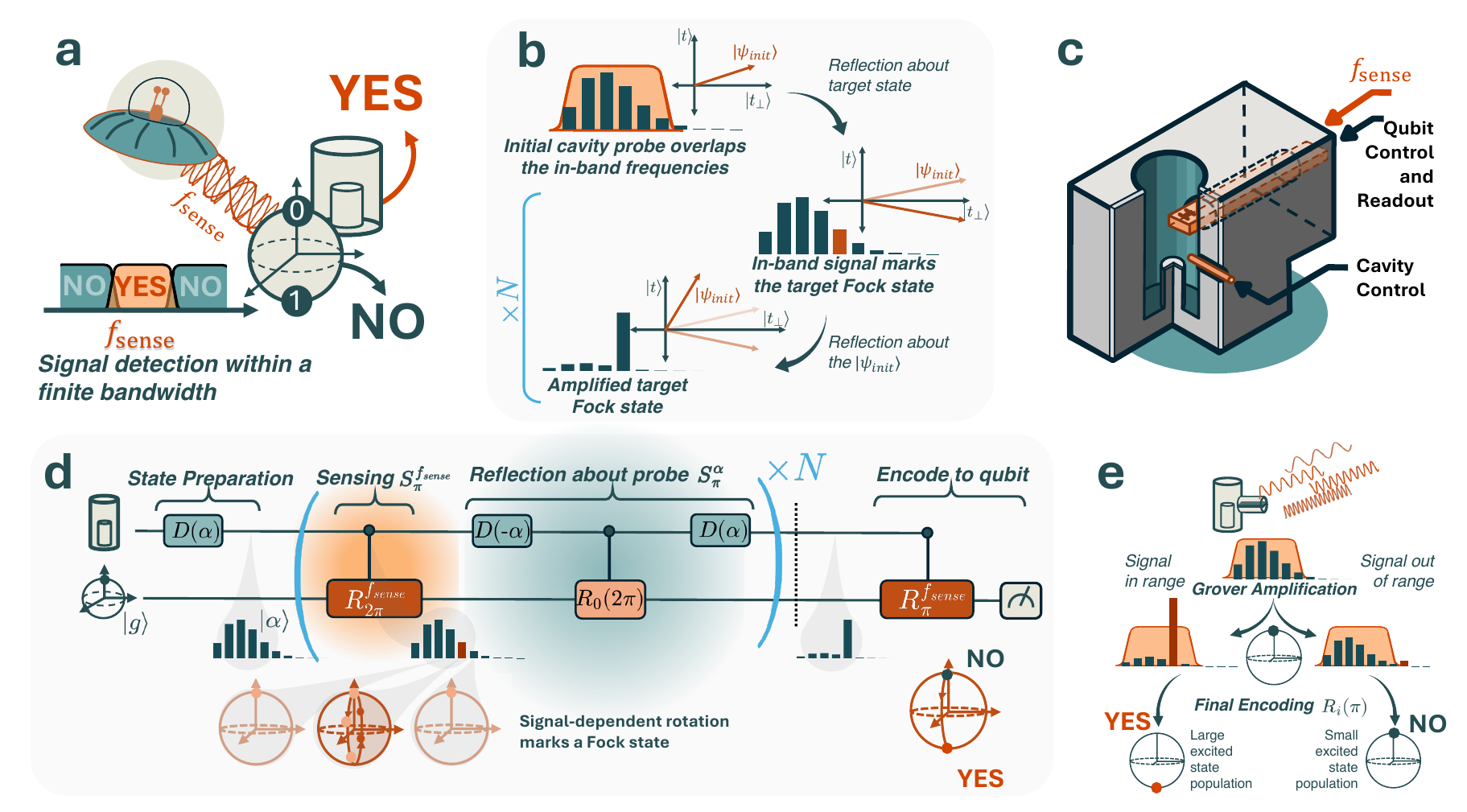}
    \caption{
    \textbf{Grover-enhanced signal detection with a superconducting cavity.} 
    \textbf{a} Binary signal classification: the sensor determines whether a signal belongs to a target frequency band.
    \textbf{b} Grover amplitude amplification in the cavity Fock-state basis. 
    Starting from a coherent probe, successive reflections about the signal-marked state and the initial probe amplify the marked-state population. 
    \textbf{c} Experimental platform consisting of a superconducting cavity dispersively coupled to a qubit. 
    \textbf{d} Pulse sequence. 
    A displacement $D(\alpha)$ prepares the coherent probe.
    Each Grover iteration combines a signal-induced rotation $R_i(2\pi)$ with a reflection about the initial probe, implemented using $D(-\alpha)$, a vacuum-selective rotation $R_0(2\pi)$, and $D(\alpha)$. 
    After $N_{\mathrm{Grover}}$ iterations, a final signal pulse $R_i(\pi)$ maps the resonant Fock-state population onto the qubit-excitation probability. 
    Histograms and Bloch spheres illustrate the cavity populations and conditional qubit dynamics. 
    \textbf{e} Classification principle. 
    Signals within the sensing window address amplified Fock-state populations and produce a large qubit-excitation probability, whereas signals outside the window produce a small response.
    }
    \label{fig:fig1}
\end{figure*}

Quantum sensing is the use of quantum systems to measure physical quantities~\cite{ Giovannetti2004, giovannetti_advances_2011,Degen2017,pirandola2018advances}. When the end goal of quantum sensing is to make a decision based on a sensed signal rather than to precisely reconstruct the signal~\cite{tsang2012continuous,martinez2021quantum}, there can be substantial advantages to combining quantum sensing with quantum computing, called quantum computational sensing~\cite{Khan2025QCSA}. In this paradigm, one performs a task-specific quantum computation in conjunction with a quantum sensor, enhancing the information that is revealed by quantum measurements to be that needed to perform the task. It was recently discovered~\cite{Allen2025QCES} that Grover's algorithm for quantum search~\cite{grover1997quantum,Zalka1999}, which gives a quadratic improvement in the number of queries required to find a marked item in an unordered database, can be profitably used in a quantum-computational-sensing scheme for detecting whether a signal appears in a particular sensing bandwidth or not by reframing the task as a search problem.

This is an exciting development because it provides a scenario where Grover's algorithm can plausibly give a practical benefit in the near term. For purely computational tasks---i.e., tasks not involving quantum sensing---such as the database search task for which Grover's algorithm was originally proposed, the demands on the quantum computer size, fidelity, and speed to obtain a practical advantage appear prohibitive, even in the fault-tolerant setting \cite{babbush2021focus}. However, in the setting of quantum computational sensing, where the relevant resource is the signal to be sensed rather than the computing time, the situation is completely different: even with a very small quantum processor, it is possible to achieve an advantage over the best possible alternative strategy \cite{Khan2025QCSA,Sen2026QCS}.

In this paper, we report a proof-of-concept experimental demonstration of the Grover-based quantum-sensing approach from Ref.~\cite{Allen2025QCES}. Our experimental platform is a single superconducting transmon qubit~\cite{krantz2019quantum} dispersively coupled to a single superconducting cavity~\cite{blais2020quantum} (Fig.~\ref{fig:fig1}c). 
We demonstrated a yes/no binary decision task (Fig.~\ref{fig:fig1}a): the same pulsed oscillating signal with an unknown center frequency is impinging on the quantum system at known intervals, and the task is to report \textit{yes} if the signal is inside a specified detection bandwidth, or \textit{no} otherwise. We have shown that by using the Grover-based approach (Fig.~\ref{fig:fig1}b,d,e), one can perform this task for large bandwidths using less sensed signal than is required to achieve the same task accurately with a conventional protocol, illustrating a Grover-based enhancement in sensing time. For small detection bandwidths ($<10$~MHz), the baseline conventional protocol performed better than the Grover-based protocol. However, at a detection bandwidth of approximately 10~MHz, there was a cross-over where the Grover-based protocol performed approximately as well as the baseline protocol, and for detection bandwidths larger than 10 MHz, the Grover-based protocol performs better, with an enhancement that increased superlinearly with the detection bandwidth.

We end the paper with a discussion, highlighting various limitations of our experiments and what potential future research might advance Grover-based sensing from our proof-of-concept results to practical benefits in real sensing use cases.

\section{Grover-based quantum sensing using superconducting circuits}
\label{sec:amp_amplification}

We first describe how the Grover-based sensing protocol is implemented in our superconducting qubit–cavity system and characterize the resulting frequency-dependent response. 
We then use this response in Sec.~\ref{sec:freq_classification} to perform binary signal classification and quantify the sensing enhancement over the coherent-probe baseline.
Our sensor encodes different signal frequencies into distinct cavity Fock states using a fixed-frequency qubit coupled to a cavity.
In the dispersive regime, photon-number splitting resolves the cavity Fock states in the qubit spectrum \cite{schuster2007resolving, Blais_CQED_rev} (See Appendix~\ref{app_sec:experimental_setup} and~\ref{app_sec:calibration} for device and calibration details). 
Each Fock state $\ket{i}$ is associated with a distinct qubit transition frequency $f_{ge,i}$, providing a set of frequency-selective sensing probes.
An incoming signal centered at $f_{ge,i}$ addresses the qubit transition associated with a specific Fock state.
We use this transition-specific interaction to drive the qubit along a closed path on the Bloch sphere to perform a $2\pi$ rotation ($R_i(2\pi)$) (see Fig.~\ref{fig:fig1}d).
This rotation implements a Selective-Number-Dependent Arbitrary Phase (SNAP) gate on the associated Fock state, set by the solid angle enclosed by the qubit trajectory on the Bloch sphere \cite{heeres2015cavity,krastanov2015universal}.
These gates serve as the oracles in our Grover-enhanced sensing protocol. 
Combined with cavity displacements, they also provide universal control of the cavity state.

We use a coherent probe state instead of an equal Fock-state superposition, simplifying amplitude amplification at the cost of a nonuniform response across the sensing range.
We prepare the coherent state by driving the cavity resonantly through its control port, implementing the unitary $D(\alpha)$ with the tone phase setting $\arg(\alpha)$  (see Fig. \ref{fig:fig1}.c).
The Fock-state populations follow a Poisson distribution,
\begin{equation}
p_i = |\langle i \vert \alpha \rangle|^2 = e^{-|\alpha|^2}\frac{|\alpha|^{2i}}{i!},
\qquad \bar{n}=|\alpha|^2.
\end{equation}
Due to the dispersive frequency shifts, this distribution sets the sensor’s frequency coverage. The initial populations ($p_i$) set how strongly the corresponding Fock component is amplified after a given number of Grover iterations.

The Grover protocol amplifies the marked Fock component by alternating reflections about the marked state $\ket{i}$ and the initial probe state $\ket{\alpha}$.
The first reflection, $S_{\pi}^{(i)}$, is implemented by a $2\pi$ qubit rotation that imparts a $\pi$ phase on the resonant Fock component.
The second reflection about the initial probe state is implemented using cavity displacements and a vacuum-selective $\pi$-phase SNAP gate $R_{0}(2\pi)$, such that
\begin{equation}
S_{\pi}^{(i)} = R_i(2\pi)= I - 2\lvert i,g\rangle\langle i,g\rvert,\qquad
S_{\pi}^{(\alpha)} = D(\alpha)R_0(2\pi)D(-\alpha)
= I - 2\lvert\alpha,g\rangle\langle\alpha,g\rvert.
\end{equation}

\begin{figure*}
    \centering
    \includegraphics[width=0.75\linewidth]{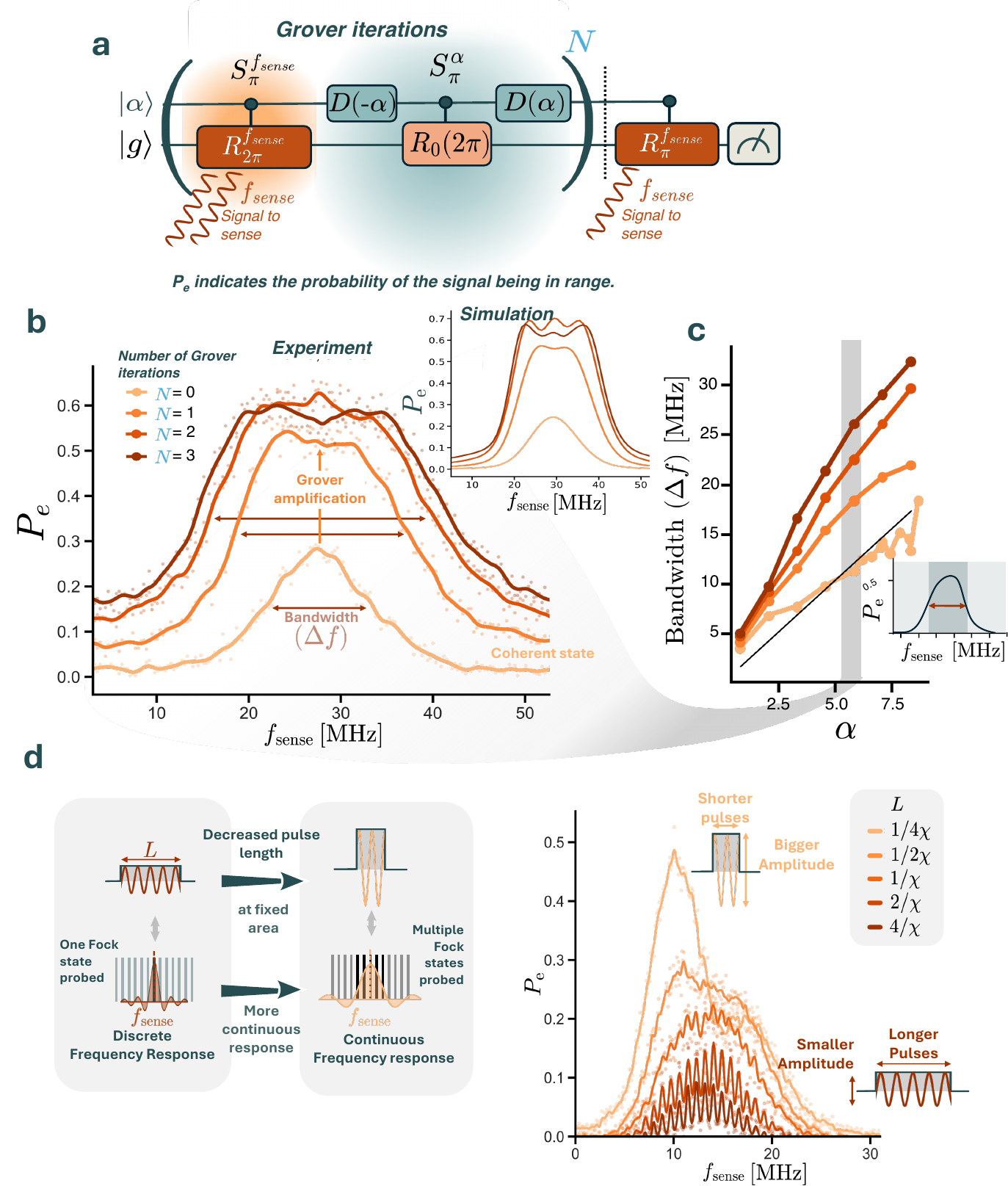}
    \caption{ 
    \textbf{Amplitude amplification using Grover's algorithm and its broadening of the bandwidth of the sensing window.}
    \textbf{a} A coherent state $\ket{\alpha=5.8}$ is subjected to $N$ Grover iterations, where a signal pulse centered at frequency $f_{\mathrm{sense}}$ produces number-selective qubit rotations.
    For the experiments shown here, the signal-pulse duration is $0.25/\chi$.
    A final signal pulse maps the classification result onto the qubit, such that the qubit-excitation probability $P_e$ indicates whether $f_{\mathrm{sense}}$ lies within the sensing window.
    \textbf{b} Measured qubit-excitation probability as a function of sensing frequency for $N=0$--$3$ Grover iterations. 
    Grover-based amplification increases both the excitation probability and the frequency range over which the signal is classified as in-band. 
    The upper inset shows the corresponding simulated responses. 
    \textbf{c} Sensing bandwidth $\Delta f$ as a function of the initial displacement amplitude $\alpha$ for different numbers of Grover iterations. 
    The solid black curve shows the theoretical predictions.
    \textbf{d} Schematic illustrating the transition from discrete to continuous sensing.
    Decreasing the signal-pulse duration at fixed pulse area decreases the spectral selectivity, changing the response from resolving individual number-split transitions to simultaneously probing multiple neighboring Fock states.
    }
    \label{fig:fig2}
\end{figure*}

The inverse displacement maps the coherent probe onto the vacuum, after which the SNAP gate applies a reflection about the vacuum state and the final displacement restores the original basis.
The resulting Grover operator,
$G=S_{\pi}^{(\alpha)}S_{\pi}^{(i)},$
implements one Grover iteration, as shown in Fig.~\ref{fig:fig2}a.
The signal at frequency $f_{\mathrm{sense}}$ implements the ideal reflection $S_{\pi}^{(i)}$ only when it is resonant with the transition associated with Fock state $i$, when $f_{\mathrm{sense}}=f_{ge,i}$. 
For a signal $f$ detuned from $f_{ge,i}$, the finite spectral width of the pulse produces a more general number-dependent phase operation, which we denote by $S_{\pi}^{(f)}$.

We first verify that the sensing signal produces Grover amplification of the addressed Fock component. 
Starting with a coherent probe state, we apply alternating reflections about the marked state and the initial probe state to increase the population of the marked Fock state, and perform photon number resolved measurements. 
The results (Appendix Fig.~\ref{fig:fock_resolved}) show that amplitude amplification encodes the signal’s center frequency in the Fock basis by increasing the marked-state population.

We convert this amplitude amplification into a binary sensing outcome using a final number-selective qubit $\pi$ pulse driven at the signal frequency as shown in Fig.~\ref{fig:fig2}a.
After $N$ Grover iterations, this pulse, $R_i(\pi)$, excites the qubit conditioned on the cavity occupying the Fock state $\ket{i}$ whose number-split transition is resonant with $f_{\mathrm{sense}}$.
For an in-band signal, the corresponding Fock component has already been amplified, resulting in a high qubit-excitation probability.
By contrast, an out-of-band signal addresses a component that has not been amplified and produces a low excitation probability.
A small number of qubit measurements can thus distinguish the two cases without reconstructing the full photon-number distribution, at the cost of one additional interaction with the signal.
Figure~\ref{fig:fig2}b shows the measured qubit-excitation probability versus signal frequency for $\alpha=5.82$ and Grover iterations $N=0,1,2~\text{and}~3$.
Under ideal amplitude amplification, the probability of occupying the addressed Fock state after $N$ iterations is
$P_N(i,\alpha)=\sin^2\left[(2N+1)\arcsin\sqrt{p_i}\right]$.
Simulations including finite pulse durations reproduce the measured response (Fig.~\ref{fig:fig2}b, inset).

The number of Grover iterations ($N$) controls how the initial Fock-state distribution is amplified.
The theory predicts the optimal $N$ such that the central Fock component reaches its first maximum, $(2N+1)\arcsin\sqrt{p_{\bar{n}}}\simeq \frac{\pi}{2}$. 
This produces an amplified response centered near $\bar{n}$ and extending over a broad range of neighboring Fock states (see Fig.~\ref{fig:fig2}b).
Increasing $N$ amplifies initially weaker components farther from the center and eventually over-rotates the center.
The ideal scaling of the optimal iteration count and sensing-window width is derived in Appendix~\ref{app_sec:theory}.

The displacement amplitude determines both the center and the width of the sensing-frequency window.
We define the sensing bandwidth $\Delta f$ as the FWHM of the measured qubit-excitation response. 
Frequencies inside this window form class 1, while neighboring out-of-band frequencies form class 0 (see Appendix~\ref{app_sec:analysis}  for more details).
The maximum response occurs near the qubit transition associated with the most highly populated Fock component of the initial coherent state.
Because its photon-number distribution is centered at $\bar{n}=|\alpha|^2$, the dispersive frequency mapping gives the center frequency $f_c\simeq \chi|\alpha|^2$ (See Appendix Fig.~\ref{fig:fock_resolved}c).
The sensing-window width scales approximately linearly with $\alpha$, with $\mathrm{FWHM}\simeq 4|\alpha|\chi\sqrt{\ln 2}$ (see Fig. \ref{fig:fig2}.c and  Appendix~\ref{app_sec:theory}).
Amplitude amplification also broadens the sensing bandwidth by enhancing the response to frequencies near the edges of the coherent probe’s sensing window, as shown in Fig.~\ref{fig:fig2}b.

Deviations from this ideal scaling arise from two effects. 
First, the finite selectivity of the number-resolved pulse ($R_i(2\pi)$) broadens the response on both sides, adding an approximately pulse-limited contribution to the sensing bandwidth.   
Second, this imperfect selectivity compounds over successive Grover iterations, introducing a weak dependence of the bandwidth on $N$.
This broadening motivates examining the signal-pulse bandwidth separately.

The signal-pulse duration controls the transition from discrete to continuous signal detection.
Long pulses spectrally resolve individual number-split transitions and therefore implement nearly selective $R_i(\pi)$ rotations, producing a comb-like response as seen in Fig.~\ref{fig:fig2}d. 
Shorter pulses partially address neighboring number-split transitions. 
Shorter pulses have a broader spectral response, allowing signals between adjacent number-split transitions to drive imperfect qubit rotations.
Although these rotations no longer correspond to ideal selective Grover operations, amplitude amplification can remain effective for non-$\pi$ rotations.
Figure~\ref{fig:fig2}d shows the qubit-excitation probability after one iteration of amplitude amplification for pulse lengths $L=1/4\chi,\,1/2\chi,\,1/\chi,\,2/\chi,$ and $4/\chi$, with $\chi/2\pi \approx 0.86~\mathrm{MHz}$.
For progressively longer pulses, we see finer spectrally resolved features, which disappear for shorter pulse lengths.
In our implementation, the overlap between these partially selective rotations fills the gaps between adjacent transitions and produces a continuous sensing window (see Fig.~\ref{fig:fig2}d).

Limited pulse selectivity also increases the measured response amplitude. 
During the final sensing step, a broader pulse maps population from several neighboring Fock states onto the qubit, increasing the excitation probability even for $N=0$, as seen in Fig.~\ref{fig:fig2}b. 
During the Grover iterations, the same finite selectivity drives partial rotations of neighboring states, which can also contribute to amplitude amplification.
Together, these controls make the Grover-amplified response a tunable frequency window that can be used for classification.

\section{Enhanced signal detection within a finite bandwidth using Grover-based sensing}
\label{sec:freq_classification}

\begin{figure*}[h!]
    \centering    \includegraphics[width=\linewidth]{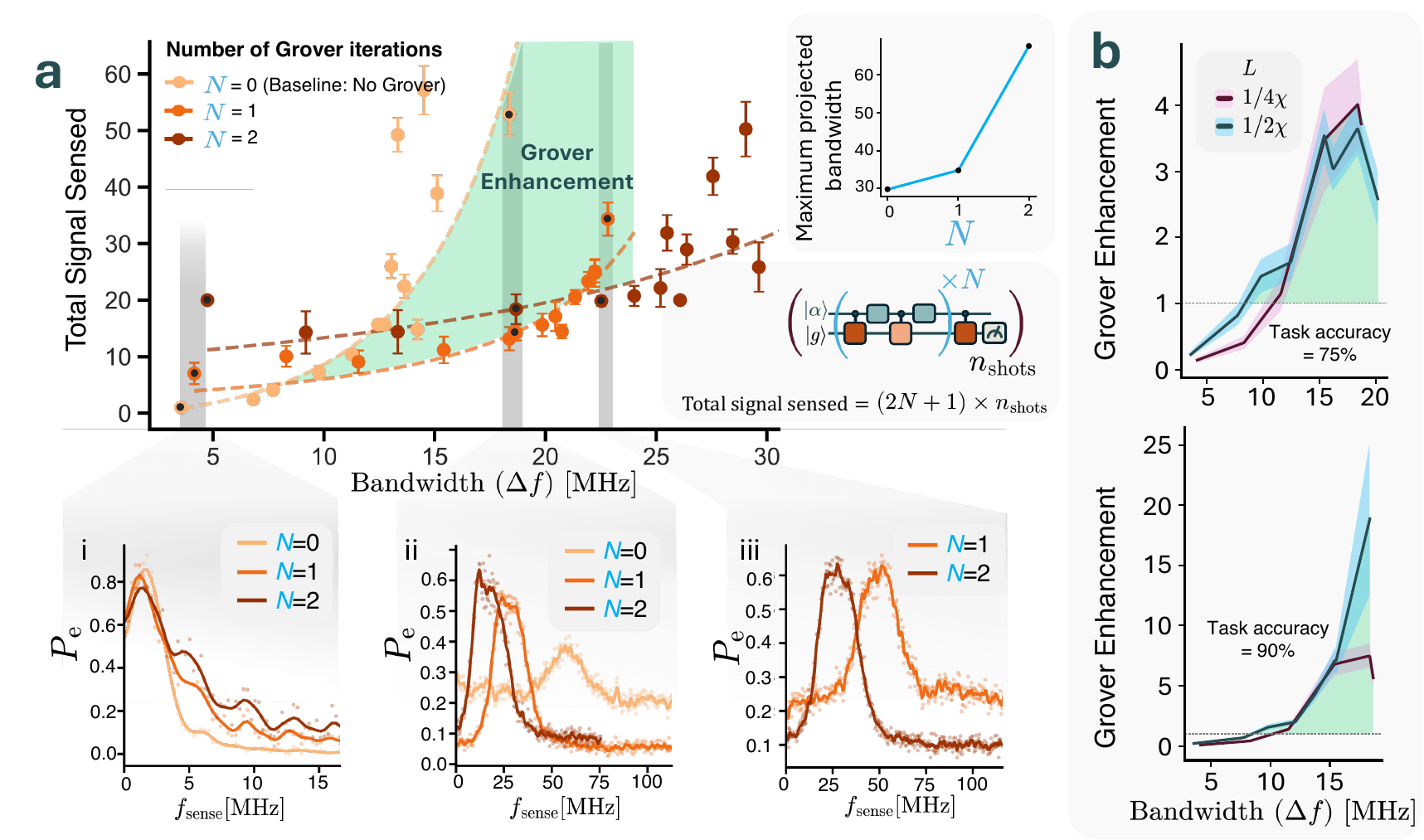}
    \caption{\textbf{Demonstration of a Grover-based enhancement in quantum sensing.} \textbf{a} Scatter plot of total signal sensed, defined as $n_{\mathrm{shots}}(2N+1)$, required to reach $75\%$ classification accuracy as a function of the target bandwidth $\Delta f$ for a signal-pulse duration of $0.25/\chi$. Classification accuracies and bandwidths were calculated for a range of initial displacements using methods detailed in Appendix~\ref{app_sec:analysis}. 
    A protocol with $N$ Grover iterations has a sensing cost of $2N+1$ signal-interaction per shot.
    Colors denote the number of Grover iterations $N$, with $N=0$ giving the coherent-probe baseline.
    Points show the mean over 100 trials, and error bars show the standard deviation across these trials. 
    In each trial, 1,000 points are randomly drawn from a dataset of 10,000 points, with 800 used for training and 200 for validation.
    Dashed curves are fits to the shot-estimate model described in the Appendix~\ref{app_sec:Fit_model}.
    The green shaded region marks the bandwidth range over which Grover-based sensing requires less total signal sensed than the coherent-state baseline.
    Gray vertical bands indicate the three representative bandwidths shown in panels \textbf{(i)}--\textbf{(iii)}.
    Grover-based sensing achieves a given sensing bandwidth at a smaller coherent-state amplitude \(\alpha\), extending the bandwidth accessible before possible breakdown of the dispersive approximation.
    The upper-right inset shows the bandwidth at which the fitted shot requirement diverges as the contrast between in-band and out-of-band signals vanishes. 
    Unlike the response bandwidth in Fig.~\ref{fig:fig2}c, this quantity characterizes the bandwidth limit imposed by the resources required for classification. 
    Its increase with the number of Grover iterations shows that amplitude amplification extends the range over which classification can be performed with finite resources.
    \textbf{b} Grover enhancement, defined as the ratio of the total signal sensed by the coherent-probe baseline to that required by the Grover protocol, as a function of sensing bandwidth for signal-pulse durations $L=1/4\chi$ and $1/2\chi$. 
    Results are shown for target classification accuracies of $75\%$ (top) and $90\%$ (bottom). 
    The horizontal dashed line marks unity, above which the Grover protocol provides a resource advantage, highlighted in green.
    }
    \label{fig:fig3}
\end{figure*}

We use the Grover-amplified response to classify whether the signal to sense lies inside or outside a frequency window of bandwidth $\Delta f$ centered at $f_c$.
The sensor was probed with signal frequencies sampled over the range of $[f_{ge,0}+\chi,f_{ge,0}-4.5\times f_c]$, from which the in-band and out-band regions are determined using the method detailed in Appendix~\ref{app_sec:analysis}.
We run our Grover-enhanced classification protocol illustrated in Fig.\ref{fig:fig3}a to map the classification information onto the qubit excitation probability, with higher excitation probability indicating an in-band signal. 
The classification accuracy accounts for both in-band detection and out-of-band rejection, and therefore includes both false-negative and false-positive errors.
We compare protocols employing one and two Grover iterations with a coherent-probe baseline by evaluating the minimum number of signal interrogations required to reach a fixed classification accuracy of $75\%$.

We compare the Grover protocol with the corresponding sensing sequence performed without amplitude amplification ($N=0$).
The baseline uses the same coherent-state preparation and final signal-dependent qubit rotation without the intermediate Grover reflections. 
We adjust the coherent-state amplitude for each protocol to compare their resource requirements at the same sensing bandwidth. 
Without the cavity, the fixed-frequency qubit can probe only a single transition and therefore cannot access a broad frequency range.
The cavity creates a set of photon-number-resolved qubit transitions, while the coherent displacement populates the corresponding Fock states.
A coherent-state probe ($N=0$) therefore provides the natural baseline for broadband sensing with this system, distributing the sensitivity across a range of frequencies.

Grover amplification trades additional signal interactions within each shot for a reduction in the number of shots required for classification.
To compare the resource requirements fairly, we define the total number of signal interrogations as $n_{\mathrm{shots}}\times (2N+1)$.
Each Grover iteration contains a signal-driven $2\pi$ rotation, counted as two interrogations, while the final $R_i(\pi)$ pulse contributes one additional interrogation.

For broad frequency windows, Grover amplification reduces the total number of signal interrogations required to reach the target accuracy.
Figure~\ref{fig:fig3} shows the total number of signal interrogations required to reach $75\%$ classification accuracy as a function of bandwidth $\Delta f$, for a signal-pulse duration of $L=0.25/\chi$.
For narrow classification windows, the coherent-probe baseline ($N=0$) reaches the target accuracy with fewer signal interrogations. 
As the bandwidth increases, however, its resource requirements grow faster than those of the Grover-enhanced protocols.
Both the $N=1$ and $N=2$ protocols cross the coherent-probe baseline at bandwidths of approximately $8~\mathrm{MHz}$ and $13~\mathrm{MHz}$, respectively, beyond which they require fewer signal interrogations.
Above approximately $22~\mathrm{MHz}$, the $N=2$ protocol also outperforms $N=1$ (see Appendix Fig.~\ref{fig:supp_theory} for comparison with the optimal theory).
Grover amplitude amplification increases the contrast between the qubit responses to in-band and out-of-band signals, allowing the $N=1$ and $N=2$ protocols to reach the target accuracy with fewer total signal interrogations at larger bandwidths. 
We repeat the analysis for a target classification accuracy of $90\%$ and observe the same qualitative trends (Appendix~\ref{app_sec:more_exp} and Fig.~\ref{fig:acc_90}).

Grover amplification extends the experimentally accessible sensing bandwidth, independent of its resource-cost advantage.
At higher Fock states, several non-idealities become increasingly important, including reduced readout fidelity, breakdown of the dispersive approximation, and imperfect number-selective rotations. 
These effects limit how far the coherent-state bandwidth can be increased simply by increasing $\alpha$.
Grover amplification provides the bandwidth enhancement at lower Fock-state occupation, before these high-photon-number limitations become dominant. 
Figure~\ref{fig:fig3}(i--iii) compares qubit responses with the same sensing bandwidth. 
Increasing the number of Grover iterations achieves a given bandwidth at smaller displacement amplitudes, demonstrating this advantage.
We use the fits to the resource requirement (dashed curves in Fig.~\ref{fig:fig3}, see Appendix~\ref{app_sec:Fit_model}) to estimate the maximum bandwidth attainable by each protocol, as shown in the upper-right inset. The maximum projected bandwidth increases from approximately $30~\mathrm{MHz}$ for $N=0$ to $35~\mathrm{MHz}$ and $68~\mathrm{MHz}$ for $N=1$ and $N=2$, respectively.

The effective-bandwidth gain from Grover amplification persists for longer, more spectrally resolved signal pulses.
We repeat the measurements and analysis for pulse durations of $0.5/\chi$ and $1/\chi$ (see  Appendix~\ref{app_sec:more_exp} and Fig.~\ref{fig:acc}).
For both pulse durations, the Grover protocols retain their resource advantage over the coherent-probe baseline, while the integrated response increases with the number of Grover iterations.
This behavior is consistent with the increase in the integrated sensing response produced by Grover amplification (see Appendix~\ref{app_sec:Fit_model}).
These results show that the advantage is not specific to the shortest signal pulse, but persists as the frequency response becomes more spectrally resolved.

To quantify the sensing enhancement obtained from Grover-based amplification, we plot the Grover enhancement, defined as the ratio of the total signal sensed by the coherent-probe baseline to that required by the Grover protocol at the same bandwidth, as shown in Fig.~\ref{fig:fig3}b. 
An enhancement larger than one therefore indicates a resource advantage. 
The enhancement increases with bandwidth, reaching approximately $4\times$ for a target accuracy of $75\%$ and up to $18\times$ for $90\%$, showing that the benefit of Grover amplification becomes more pronounced for broader sensing windows and higher target accuracies.

\section{Discussion}
\label{sec:discussion}

\textit{Summary:}~We demonstrated enhanced detection of a signal with an unknown center frequency by using a protocol based on Grover's algorithm for quantum search implemented in a superconducting qubit-cavity system. The Grover-based protocol was implemented by having the signal to be sensed drive (or not) qubit transitions associated with different cavity Fock states. Signals with frequencies corresponding to these transitions marked the corresponding Fock states by implementing a $2\pi$ rotation to impart a $\pi$-phase shift on the marked Fock state. We quantified the enhancement from the Grover-based protocol over a conventional baseline protocol by calculating the amount of signal that would need to be sensed to achieve a desired accuracy. Below a detection bandwidth of approximately 10~MHz, the baseline protocol performed better than the Grover-based protocol, but beyond 10~MHz we observed an advantage from the Grover-based protocol that grew superlinearly with bandwidth. For 90\% accuracy, the Grover-based protocol needed as much as $18\times$ less signal than the baseline protocol.

A key aspect of our work is achieving a continuous detection bandwidth. Since the qubit-cavity transition spectrum is discrete, to detect signals over a continuous window we used pulses with spectral width that was comparable to the frequency spacings in the qubit-cavity spectrum. This allows frequencies between two adjacent transitions to drive imperfect qubit rotations. We found that Grover's amplitude amplification remains robust and able to provide an enhancement under these conditions~\cite{Grover1998}.

\textit{Limitations:}~Our study has various limitations that constrain how our results should be interpreted and how readily our work could be applied to practical sensing scenarios. The following list is not exhaustive---and most points apply not only to our work, but to works within the fields of quantum sensing and quantum computational sensing more generally~\cite{Khan2025QCSA}. First, as is the case in the theoretical proposal~\cite{Allen2025QCES}, the protocol we implemented assumes that the signal arrives at known times and that we can synchronize our system with it. Second, we assumed that the signal repeats as many times as we need to make a high-accuracy detection decision---but in practice many signals are transient. Third, we evaluated the Grover-based protocol against a baseline protocol that we have argued is natural as a conventional sensing approach but we have not proven that this baseline protocol is the best possible non-Grover-based protocol for our experimental platform. Fourth, instead of preparing a uniform superposition of Fock states as the initial state for the protocol, and performing rotations about that state, we prepared a coherent state. 
This makes the protocol more sensitive to some signal frequencies and less sensitive to others~\cite{Biham1999,Brassard2000};
the Grover enhancement could likely be even greater if a uniform superposition were used instead. Fifth, the bandwidth over which we performed sensing was rather modest---up to tens of MHz. Sixth, we only demonstrated up to two Grover iterations because we were limited by the coherence time of our qubit; more iterations could allow greater detection bandwidth and even larger advantage. Seventh, the signal to sense was generated at room temperature but sent through various stages of attenuation before impinging on the qubit-cavity system, reducing the absolute sensitivity of the sensing system with respect to the original room-temperature signal (equally for both the Grover-based and baseline protocols). Use cases where signals arise at the cryogenic operating temperature for our superconducting system of tens of millikelvin---for example, in dark-matter searches~\cite{backes2021quantum}---might directly be able to take advantage of Grover-based sensing without using any attenuation, but it is an open question whether sensing of signals that originate at room temperature is practical with the qubit-cavity system we used.

\textit{Outlook:}~Each of the limitations we have outlined suggests a possible future direction to eliminate or otherwise mitigate it. For example, it may be possible to engineer uniform Fock-state superpositions instead of using the coherent-state approximation~\cite{hofheinz2009synthesizing, wang2017converting}. Another interesting direction for future work that could be enabled by Grover's algorithm is, instead of addressing the yes/no detection task we studied, to tackle the task of identifying which discrete frequency bin a signal falls in. Yet another would be to implement the quantum-signal-processing~\cite{LowYoderChuang2016,LowChuang2017,MartynEtAl2021} step in the original theoretical proposal~\cite{Allen2025QCES} to discretize the phases that accrue during the sensing operations. One could also attempt to compute functions of the frequency of the sensed signal beyond the simple rectangular function used in the binary task we studied~\cite{Sen2026QCS}, or to study the equivalent of the multiple-marked-items case in Grover's algorithm: when signals comprising multiple frequencies arrive simultaneously ~\cite{boyer1998tight}.

Our work makes progress in addressing the longstanding challenge to find applications for quantum computing technology that are practical in near-term use cases. By repurposing quantum computing hardware and algorithms for sensing tasks, we can gain proof-of-principle advantages in problems using far lower hardware requirements than are needed for purely computational tasks. Our results provide experimental validation of this approach, and motivate further developments to take on the challenge of achieving Grover-enhanced sensing in practical scenarios.

\vspace{0.5cm}
\textit{Note added in proof:}~During the preparation of this manuscript, we became aware of a closely related recent work, Ref.~\cite{Xu2026ExperimentalGroverSensing}, on Grover-enhanced quantum sensing using superconducting circuits. Ref.~\cite{Xu2026ExperimentalGroverSensing} presents mutual information and number of resolved discrete frequency bins as metrics for advantage, whereas in our work we present sensing time to achieve high classification accuracy on a yes/no task, like the one defined in the original theoretical proposal~\cite{Allen2025QCES}, as a function of the continuous bandwidth that the system is detecting a signal in.

\section*{Data and code availability}
All experimental and simulation data and code used in this work are available at  \url{https://doi.org/10.5281/zenodo.23000895}.

\section*{Acknowledgements}
This work was supported by the Air Force Office of Scientific Research under Award No. FA9550-22-1-0203, and the Army Research Office under Award No. W911NF-25-1-0261. M.O. acknowledges support from the Fonds de recherche du Québec Postdoctoral Research Scholarship, Grant No. 366727. The authors thank NTT Research for its financial and technical support. The authors would like to thank Saeed~Khan, Mandar~Sohoni, and Logan~Wright for the helpful discussions and comments, and Soonwon~Choi for suggesting the use of coherent states in the Grover-based protocol. The authors would also like to thank Bradley Cole, Clayton Larson, Britton Plourde, Eric Yelton, and Luojia Zhang for the fabrication of the transmon and on-chip resonator, Chris Wang for the design of the transmon, the on-chip resonator and the 3D superconducting cavity, and Nord Quantique for the fabrication of the 3D superconducting cavity. We gratefully acknowledge MIT Lincoln Laboratory for supplying the Josephson traveling-wave parametric amplifier (TWPA)~\cite{macklin2015near} used in our experiments.

\section*{Author contributions}

M.O. and P.S. designed and carried out the experiments. M.O. performed the numerical simulations. 
X.W., S.R., X.S., V.K. and S.P. contributed to the experiments. 
V.F. oversaw the design and fabrication of the superconducting devices, which were characterized and calibrated by P.S. and M.O. with help from S.R.
M.O., P.S. and P.L.M. wrote the manuscript with input from all authors. 
P.L.M. supervised the project.

\bibliographystyle{mcmahonlab}
\bibliography{references}

\appendix
\setcounter{figure}{0}
\renewcommand{\thefigure}{A\arabic{figure}}

\section{Experimental Setup}
\label{app_sec:experimental_setup}
\paragraph{Device description and system Hamiltonian} The system comprises a fixed-frequency transmon qubit coupled to a three-dimensional high-purity aluminum cavity, which acts as our storage mode and provides the computational subspace for the Grover algorithm. The transmon, made of Niobium, is fabricated on a resistive silicon chip, along with an on-chip readout resonator also made of Niobium. In the dispersive coupling regime, the dynamics of our system can be described by the Hamiltonian:
\begin{equation}
\frac{\hat{H}}{\hbar}
=
\omega_c \hat{a}^{\dagger}\hat{a}
+
\omega_q \ket{e}\bra{e}
-
\chi \hat{a}^{\dagger}\hat{a}\,\ket{e}\bra{e}
-
K\hat{a}^{\dagger 2}\hat{a}^{2}.
\end{equation}
where $\omega_q$ and $\omega_c$ are the qubit and cavity transition frequencies, $\hat{a}$ is the annihilation operator in the storage cavity Hilbert space, $\ket{e}$ is the excited state of the qubit, $\chi$ is the strength of the cross-Kerr interaction between the cavity and the qubit, $K$ is the strength of the self-Kerr interaction in the cavity, and $\hbar$ is the reduced Planck's constant. We detail the process of characterizing these terms of the Hamiltonian in section~\ref{app_sec:calibration} and have their values listed in Table~\ref{app_tab:calibration}.

\begin{figure*}[!htbp]
    
    \centering
    \includegraphics[width=0.95\linewidth]{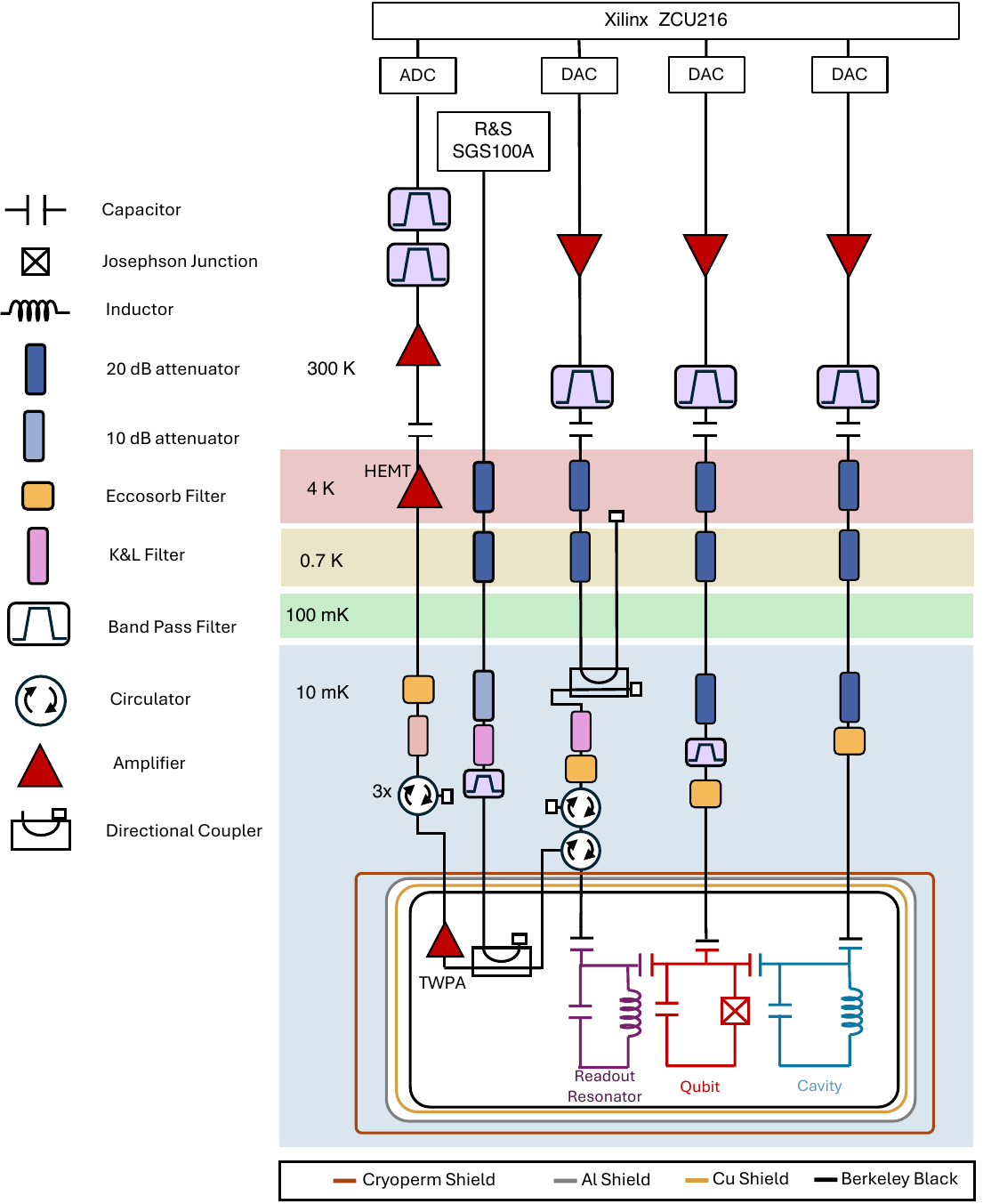}
    \caption{\textbf{Wiring Diagram:} Experimental setup for control hardware, cable routing, and shielding for our device.}
    \label{app_fig:wiring_diagram}
\end{figure*}

\paragraph{Device wiring and fridge setup}
Our device is mounted on the cold finger of a Bluefors dilution refrigerator and is shielded from radiation by a copper shield around the mixing chamber, coated with Berkeley Black on the inside. This shield is enclosed by an aluminium shield and a $\mu$-metal shield for protection from external electromagnetic fields, as shown in Fig.~\ref{app_fig:wiring_diagram}. We use a \textit{Xilinx ZCU216} board for microwave control and readout. The readout, cavity, and qubit control signals are amplified and filtered through bandpass filters and DC blocks at room temperature. The qubit and cavity control lines have 60~\si{dB} of attenuation, with 20~\si{dB} at each of the 4~\si{K}, still, and mixing-chamber (MXC) plates, and are filtered with Eccosorb filters at the MXC. The readout line is attenuated by 20~\si{dB} at each of the 4~\si{K} and still plates, with a directional coupler providing an additional 20~\si{dB} of attenuation, followed by another 10~\si{dB} of attenuation at MXC. The readout line is filtered by an Eccosorb filter and a K\&L filter, as shown in Fig.~\ref{app_fig:wiring_diagram}. The output signal is amplified by a travelling-wave parametric amplifier (TWPA) at the MXC and a high-electron-mobility transistor (HEMT) amplifier at 4~\si{K} inside the dilution refrigerator. This output is further amplified by a room-temperature amplifier before being measured by our microwave control setup. The TWPA is pumped by a microwave tone from a \textit{Rohde \& Schwarz SGS100A}, with the pump line having 50~\si{dB} of attenuation, as shown in Fig.~\ref{app_fig:wiring_diagram}.
\section{Calibration}
\label{app_sec:calibration}

\begin{table}[!htbp]

\centering
\renewcommand{\arraystretch}{1.15}
\setlength{\tabcolsep}{10pt}

\begin{tabular}{c|c|c!{\hspace{2pt}\vrule\hspace{2pt}}c}
\multicolumn{1}{c}{Parameter} & \multicolumn{1}{c}{Mode(s)} & \multicolumn{1}{c}{Symbol} & \multicolumn{1}{c}{Value} \\
\hline
\hline
Center frequency & Readout & $\omega_r$ & $2\pi\times 8846.60~\mathrm{MHz}$ \\
          & Transmon g-e     & $\omega_q$ & $2\pi\times 6266.86~\mathrm{MHz}$ \\
         & Cavity      & $\omega_c$ & $2\pi\times 7045.53~\mathrm{MHz}$ \\
\hline
Relaxation time & Transmon g-e & $T_1$   & 51~\si{\micro\second} \\

Dephasing time & Transmon g-e & $T_2^*$ & 24~\si{\micro\second} \\

Relaxation time & Cavity & $T_{1,c}$   & 85.8~\si{\micro\second} \\
Readout Error & Ground state readout error & $\mathrm{P_{eg}}$ & $5\%$ \\
                   & Excited state readout error   & $\mathrm{P_{ge}}$ & $10\%$ \\
                   \hline
Cross Kerr & Cavity-Transmon & $\chi$ & $2\pi\times0.8607$~\si{MHz}\\
Self Kerr & Cavity & $K$ & $2\pi\times1.22$~\si{kHz}\\
\hline
Ratio of gain to Xilinx amplitude & Cavity & &9.472
\end{tabular}
\caption{\textbf{Appendix Table 1 | System parameters and dissipation rates.}
}
\label{app_tab:calibration}
\end{table}
\paragraph{Calibration of frequencies, qubit coherence times, and readout fidelity}
We determine the readout frequency of the resonator by applying a weak readout pulse, sweeping its frequency, and recording its averaged IQ response. The power and frequency of the TWPA pump are adjusted to achieve a high signal-to-noise ratio. Once the frequency corresponding to the readout resonator is determined, we perform two-tone spectroscopy by keeping the readout drive frequency fixed and sweeping the frequency of the qubit drive pulse to observe a change in the averaged IQ response. When the qubit drive pulse excites the transmon, it creates a shift in the resonance frequency of the readout resonator, which is observed as a change in the IQ response on the readout line. We use this effect to drive the qubit on resonance and observe Rabi oscillations, which we use to characterize our qubit $\pi$ pulses. Readout assignment is then performed by first rotating the measured IQ data so that the separation between the $|g\rangle$ and $|e\rangle$ responses lies along a single quadrature, and then applying a threshold on this rotated axis to classify each single-shot measurement.

For determining $T_1$, we excite the qubit and measure the IQ response as a function of the delay after excitation. We fit an exponential decay curve to this IQ response to extract the decay rate and $T_1$. We also measure the Ramsey coherence time $T_2^\star$ using two detuned $\pi/2$ pulses, driven off-resonance separated by a variable delay. We fit a sinusoid with an exponentially decaying envelope to this IQ response to extract $T_2^\star$. For determining the cavity frequency, we calibrate a number-selective qubit $\pi$ pulse for the vacuum state and sweep the frequency at which we drive the cavity port. The cavity drive is followed by the vacuum-selective $\pi$ pulse, and we measure the resulting IQ response. When the cavity drive is on resonance, it prepares a coherent state in the cavity, reducing the vacuum population and hence the qubit-excitation probability after the vacuum-selective pulse. This produces a corresponding change in the measured IQ response.

\begin{figure*}
    \centering
    \includegraphics[width=0.95\linewidth]{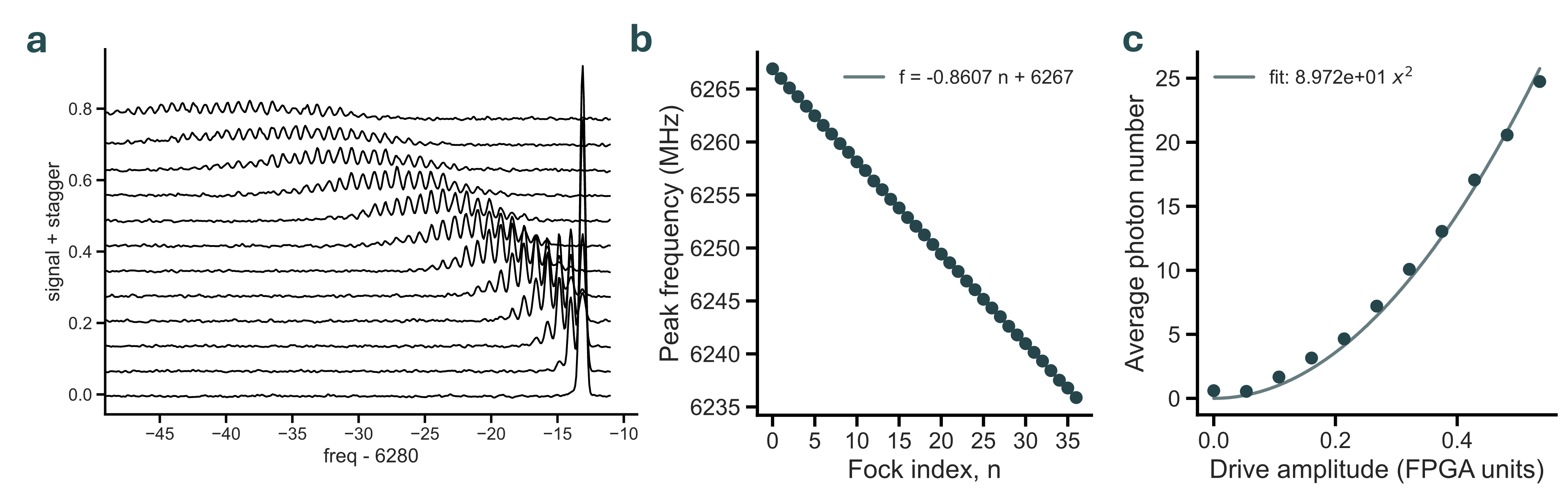}
    \caption{\textbf{Calibration of the dispersive shift and displacement amplitude.} (a) Number-splitting spectra at increasing cavity-drive amplitudes, vertically offset for clarity. (b) Qubit-transition frequencies versus Fock-state index. The linear fit gives a frequency shift of $0.8607$ MHz per photon. (c) Mean photon number extracted from Poisson fits to the measured peak heights versus drive amplitude $g$. The quadratic fit $\bar{n}=Ag^2$, with $A=89.72$, determines the displacement calibration $|\alpha|=\sqrt{A}|g|$.
    }
    \label{fig:supp_split}
\end{figure*}

\paragraph{Dispersive shift and displacement calibration}
We calibrate the dispersive shift and displacement amplitude using number-splitting spectroscopy at several cavity-drive amplitudes. 
After correcting for readout errors, we identify the resolved qubit-transition peaks and assign them consecutive Fock-state indices $n$. 
We fit their frequencies to $f_n=f_0-n\chi$, where $\chi$ denotes the frequency shift per photon (see Fig. \ref{fig:supp_split}).

For each drive amplitude $g$, we extract the peak heights and normalize their sum to obtain an estimate of the photon-number distribution. 
We fit this distribution to a Poisson model normalized over the detected Fock states to estimate the mean photon number $\bar{n}$. 
We then fit $\bar{n}=Ag^2$. 
The fitted coefficient $A$ converts the programmed drive amplitude into the coherent-state displacement through $|\alpha|=\sqrt{A}|g|$ (see Fig. \ref{fig:supp_split}).
\\

\paragraph{Cavity decay and Kerr characterization}
We characterize the cavity dynamics using time-resolved Wigner tomography for two initial displacement amplitudes, with the qubit prepared in either $\ket{g}$ or $\ket{e}$. 
We fit each distribution to a two-dimensional Gaussian in radial and angular coordinates to extract its center and widths.

\begin{figure*}
    \centering
    \includegraphics[width=0.7\linewidth]{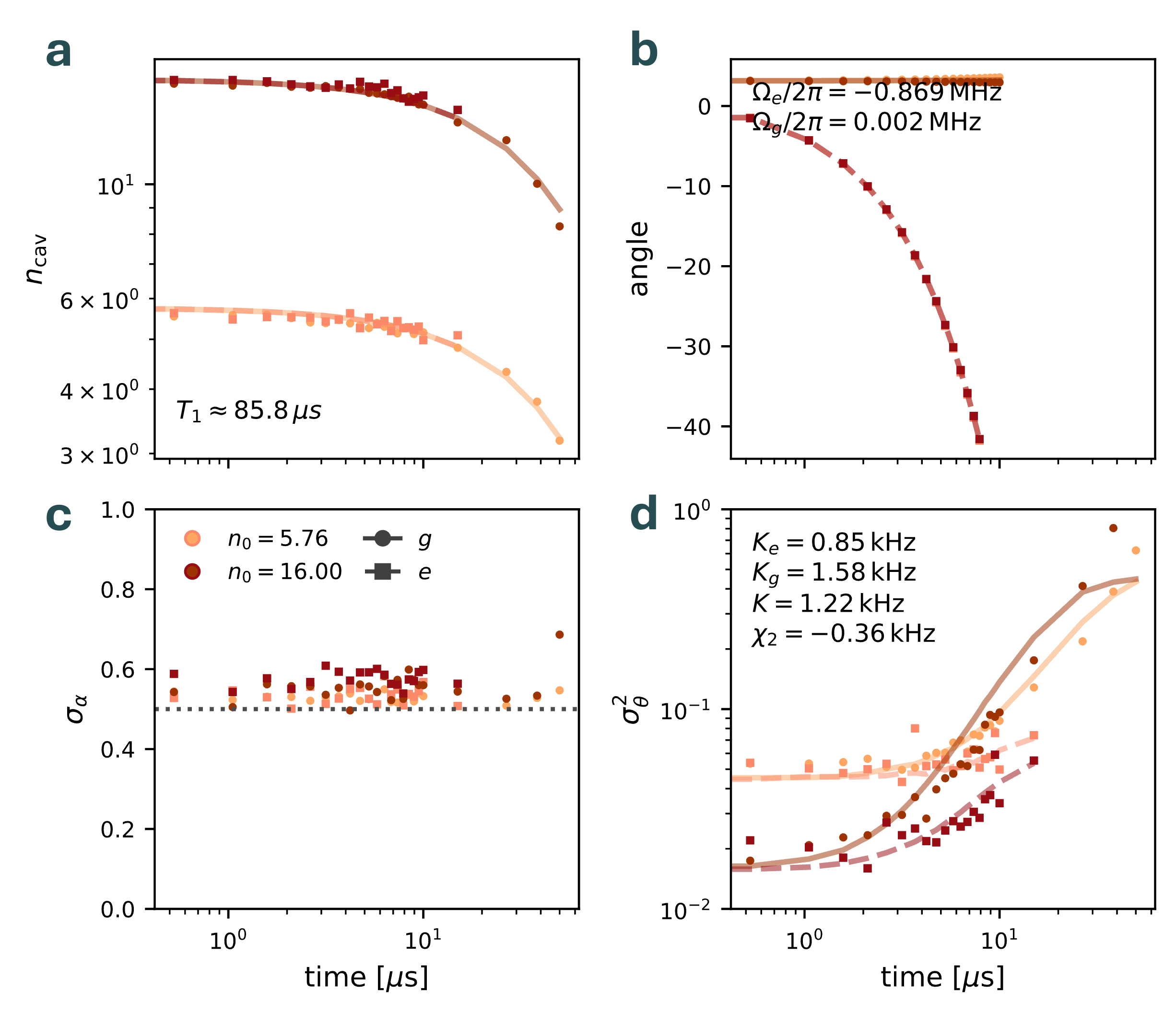}
    \caption{\textbf{
    Cavity decay and phase evolution.} Time evolution of the mean photon number (a), phase-space angle (b), radial width $\sigma_r$ (c), and angular variance $\sigma_\theta^2$ (d), for initial mean photon numbers $n_0=5.76$ and $16.00$. Circles and squares correspond to the qubit prepared in $\ket{g}$ and $\ket{e}$, respectively. 
    Curves show fits, with the extracted decay time, rotation frequencies, and nonlinear parameters indicated in the panels. 
    The dotted line marks $\sigma_r=0.5$.}
    \label{fig:supp_blob}
\end{figure*}

We estimate the photon number from the squared radial center and fit its decay to $n(t)=n_0e^{-t/T_1}$ to obtain the cavity lifetime $T_1$. 
We extract the rotation frequency for each qubit state from a linear fit to the unwrapped angular center (see Fig. \ref{fig:supp_blob}).

We estimate the effective Kerr magnitude for each qubit state by fitting the angular broadening. 
The model includes the coherent-state angular uncertainty and a Kerr-induced contribution that grows quadratically at short times, with a phenomenological saturation term at longer times.

\section{Data analysis}
\label{app_sec:analysis}

\paragraph{Measurement data loading}
We load the binary measurement outcomes for each displacement amplitude and Grover iteration count. 
For each signal frequency, we average these outcomes to obtain the qubit-excitation probability shown in the plots. 
We retain the individual outcomes for the resampling procedure used to estimate classification accuracy and the required measurement count.
\\

\paragraph{Bandwidth estimation and class assignment}
For each measured curve, we identify the class-1 frequency window automatically. 
We first smooth the curve with a Gaussian filter of standard deviation $\sigma=6$ bins and estimate the peak height as the mean of the five highest smoothed values. 
We initially estimate the baseline as the 10th percentile of the smoothed curve and set the threshold halfway between the baseline and peak height. 
Starting at the peak bin, we extend the window in both directions until the curve falls below this threshold.
We retain only curves with an SNR greater than five, defined as the difference between the peak and baseline probabilities divided by the noise level, where the noise level is estimated as the standard deviation of the difference between the raw and smoothed curves.

We then recompute the baseline using the background bins immediately flanking the initial window and repeat the procedure to obtain the final window center $m$ and width $w$.

The class-1 window contains the bins from $m-\lfloor w/2\rfloor$ to $m+\lfloor w/2\rfloor$, inclusive.
Class 0 consists of two adjacent flanking regions, each with a nominal width of $w$ bins. 
If an array boundary truncates one flank, we extend the opposite flank to compensate whenever sufficient bins are available.
These two regions define the classes used in the classification task.

\paragraph{Minimum measurement count estimation}
For each classification window, we estimate the minimum number of measurements $N_{\mathrm{measurement}}$ required to reach the target accuracy $\tau = 75 \%$. 
We first double $N_{\mathrm{measurement}}$ until the estimated accuracy exceeds $\tau$, then refine the estimate using a binary search. 
If the threshold is not reached within the maximum allowed measurement count, we report that limit and mark the estimate as unresolved. 
We repeat the search independently $N_{\mathrm{repeats}}=100$ times and report the mean and standard deviation of the resulting measurement counts.
To account for the sensing cost, we multiply these counts by $2N_{\mathrm{grover}}+1$. 
This factor includes two sensing pulses per Grover iteration and one for the final readout, giving the total number of sensing-pulse applications.

We compute accuracy using synthetic experiments on real experimental data.
For each synthetic experiment, we select one frequency bin uniformly from the assigned class. 
We sample $N_{\mathrm{measurement}}$ measured binary outcomes from that frequency bin with replacement.
We sum these outcomes to obtain the number of excited outcomes. 
We train a logistic regression classifier on the excitation counts from 80\% of the synthetic experiments and evaluate its accuracy on the remaining 20\%, preserving the class proportions in both subsets.
\\

\section{Additional Measurements and Analysis}
\label{app_sec:more_exp}

\begin{figure*}
    \centering
    \includegraphics[width=0.95\linewidth]{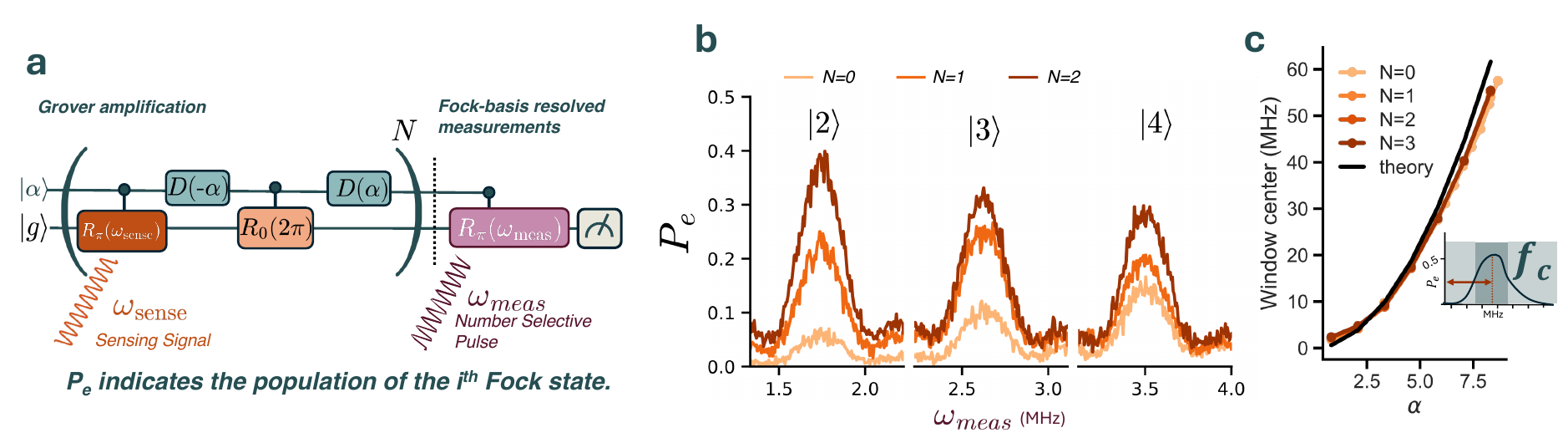}
    \caption{\textbf{Fock-state-resolved characterization of Grover amplification.}
    \textbf{(a)} Experimental sequence used to measure the population of individual Fock states after $N$ Grover iterations, starting with an initial probe, $\ket{\alpha=2.5}$. 
    Following the Grover sequence, a number-selective pulse at $f_{\mathrm{meas}}$ maps the population of the selected Fock state onto the qubit-excitation probability $P_e$.
    \textbf{(b)} Measured excitation probability as a function of the number-selective measurement frequency for $N=0$--$2$, showing amplification of the populations of the $\ket{2}$, $\ket{3}$, and $\ket{4}$ Fock states.
    \textbf{(c)} Center frequency $f_c$ of the amplified Fock-state window as a function of the initial displacement amplitude $\alpha$ for different numbers of Grover iterations. 
    The black curve shows the theoretical prediction.}
    \label{fig:fock_resolved}
\end{figure*}

\paragraph{Fock state resolved measurement:}
We prepare a coherent probe with $\alpha=2.5$ and vary the sensing frequency while keeping the reflection about the initial state fixed. After $N$ Grover iterations, we read out the population of the corresponding fock state using a number-selective qubit $\pi$ pulse at $f_{\mathrm{meas}}=f_{ge,j}$ [Fig.~\ref{fig:fock_resolved}a]. The spectra in Fig.~\ref{fig:fock_resolved}b correspond to sensing frequencies targeting $\ket{2}$, $\ket{3}$, and $\ket{4}$, from left to right. In each case, the population of the targeted state increases over the first two Grover iterations.

\begin{figure*}
    \centering
    \includegraphics[width=0.95\linewidth]{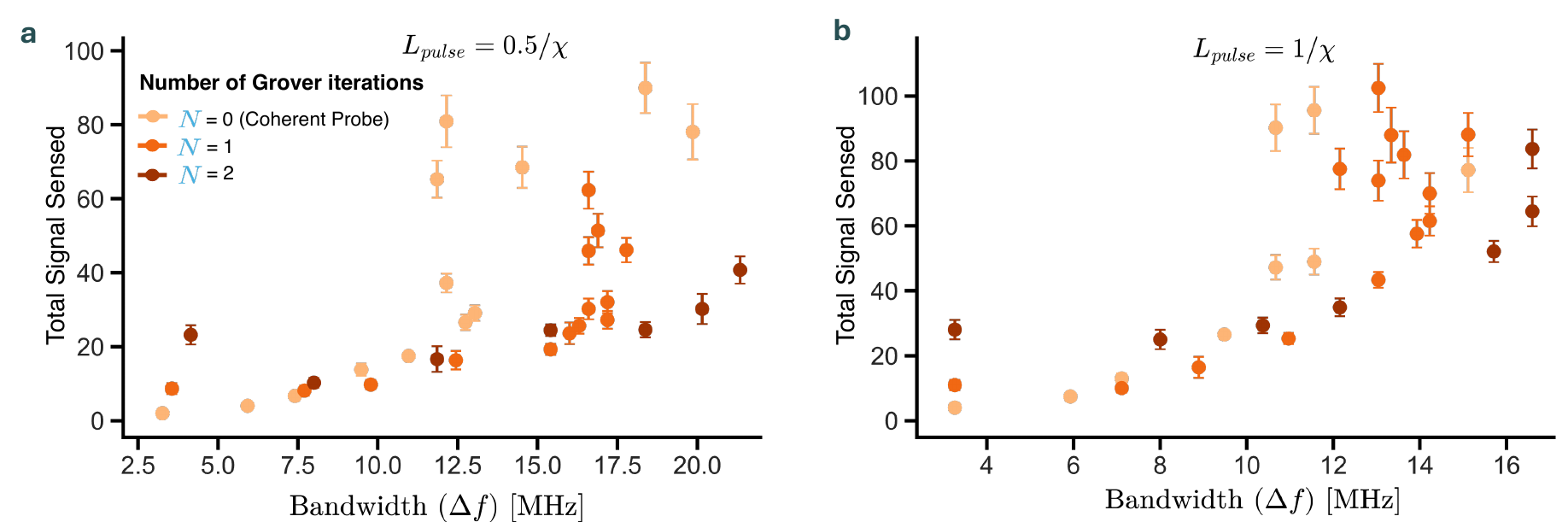}
    \caption{\textbf{Signal resources required for frequency classification at different pulse durations.} Minimum total signal required to achieve $75\%$ classification accuracy versus sensing bandwidth for sensing-pulse durations (a) $L_{\mathrm{pulse}}=0.5/\chi$ and (b) $L_{\mathrm{pulse}}=1/\chi$. Colors indicate the number of Grover iterations.}
    \label{fig:acc}
\end{figure*}

\paragraph{Frequency Classification using longer pulses:} 
We repeat the experiment detailed in \ref{sec:freq_classification}, to calculate the minimum total sensing signal required to reach a classification accuracy of $75\%$, with sensing pulse lengths of $1/2\chi$ and $1/\chi$. In both of these experiments we observe that amplitude amplified protocols provide an advantage on both classification accuracy and sensing bandwidth. 
Fig.~\ref{fig:acc} plots required signal resources versus bandwidth

\begin{figure*}
    \centering
    \includegraphics[width=0.95\linewidth]{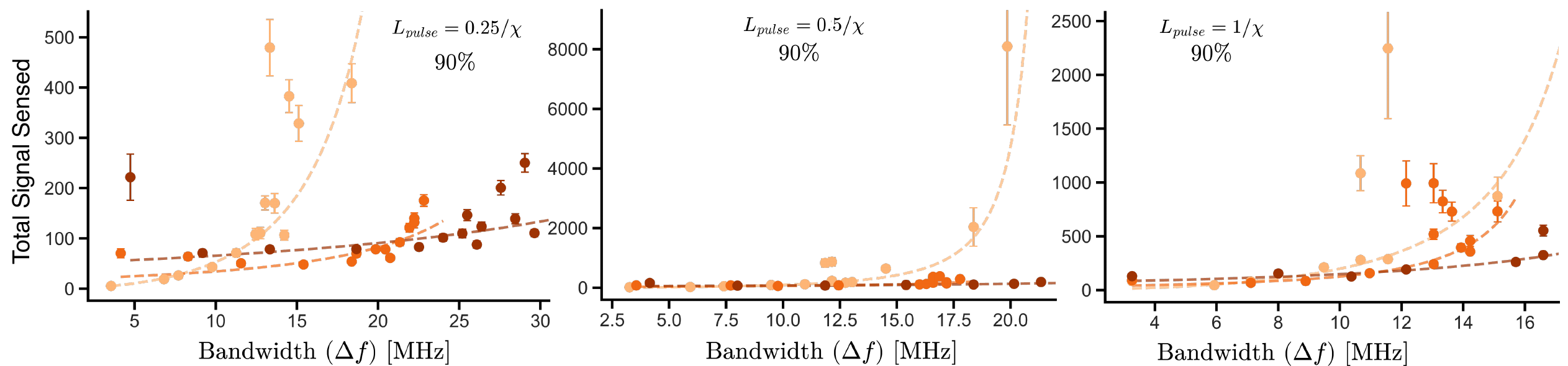}
    \caption{\textbf{Signal resources required to reach  classification accuracy of $90\%$.} Minimum signal required to achieve $90\%$ classification accuracy versus sensing bandwidth for $L_{\mathrm{pulse}}=0.25/\chi$, $0.5/\chi$, and $1/\chi$. Colors indicate the number of Grover iterations, and dashed curves show fits to the model.}
    \label{fig:acc_90}
\end{figure*}

\paragraph{Resource Requirements to achieve higher classification accuracies:} We repeat our analysis for all three pulse durations and calculate the minimum resource requirements to reach a classification accuracy of $90\%$ as shown in Fig.~\ref{fig:acc_90}. We observe that the resource advantage for Grover enhanced protocol persists, with progressively higher bandwidths favoring protocols with higher number of Grover iterations.

\section{Theoretical Analysis}
\label{app_sec:Fit_model}

\subsection{Modeling the Required Number of Shots}

We first consider the simple case of distinguishing two Bernoulli distributions with probabilities $p_l$ and $p_h$, where $p_h>p_l$. For sufficiently large $N_{\rm shot}$, the binomial distribution can be approximated by a Gaussian. Using a decision threshold between $p_l$ and $p_h$, the number of measurements required to reach a target classification accuracy $A$ therefore scales as
\begin{equation}
N_{\rm shot} \propto
\frac{z_A^2}{(p_h-p_l)^2},
\end{equation}
where $z_A=\Phi^{-1}(A)$ is the standard-normal quantile associated with the target accuracy.
For a coherent-state probe, the photon-number distribution is Poissonian such that  $p_h \sim \frac{1}{\sqrt{2\pi},|\alpha|}$.
Even in the ideal case of zero background, $p_l=0$, the required number of measurements scales as $N_{\rm shot}\propto \frac{1}{p_h^2}\propto |\alpha|^2$.

In practice, however, the out-of-band excitation probability remains finite, $p_l>0$. 
Experimentally, we observe an approximately linear increase of this background with coherent-state amplitude, $p_l=c|\alpha|$. 
The required number of measurements therefore scales as
\begin{equation}
    N_{\rm shot}
    \propto
    \frac{|\alpha|^2}{
    \left(
    \frac{1}{\sqrt{2\pi}}
    -c|\alpha|^2
    \right)^2
    }.
\end{equation}
Thus, while the ideal measurement cost scales as $|\alpha|^2$, a finite background leads to an even faster increase with coherent-state amplitude.

In addition, the finite sensing-pulse bandwidth modifies the measured peak excitation probability from the ideal coherent-state value, which we model as $p_h \simeq b/|\alpha|$. This gives
\begin{equation}
N_{\rm shot}
=
A\frac{|\alpha|^2}
{\left(b-c|\alpha|^2\right)^2},
\end{equation}
where $A$ sets the overall measurement-cost scale, $b$ captures the effective peak height, including finite pulse-bandwidth effects, and $c$ describes the increase of the background with coherent-state amplitude.

For $N>0$, Grover amplification changes the dependence of the in-band probability on $|\alpha|$. 
In the ideal case, the marked-state population after amplification can remain approximately constant as $|\alpha|$ increases, instead of decreasing as $1/|\alpha|$ as for the coherent-state baseline. 
The measurement cost is therefore no longer dominated by the decreasing peak height, but mainly by the increasing out-of-band background. 
Taking $p_h\simeq b$ we obtain
\begin{equation}
N_{\rm shot}
=
\frac{A}{\left(b-c|\alpha|\right)^2}.
\end{equation}

Although these expressions contain three parameters, only two independent
combinations enter the fit. Factoring out $c$ and defining an overall scale
$A'=A/c^2$ and a divergence point $\alpha_{\rm pole}$ reduces the models to
\begin{equation}
N_{\rm shot}
=
A'
\frac{|\alpha|^2}
{\left(\alpha_{\rm pole}^2-|\alpha|^2\right)^2},
\qquad N=0,
\end{equation}
and
\begin{equation}
N_{\rm shot}
=
\frac{A'}
{\left(\alpha_{\rm pole}-|\alpha|\right)^2},
\qquad N>0.
\end{equation}
We therefore fit the experimental data using only two free parameters: the
overall scale $A'$ and the divergence point $\alpha_{\rm pole}$.
\\

\textbf{Quantifying the gain from Grover iterations:}
\begin{figure*}
    \centering
    \includegraphics[width=0.95\linewidth]{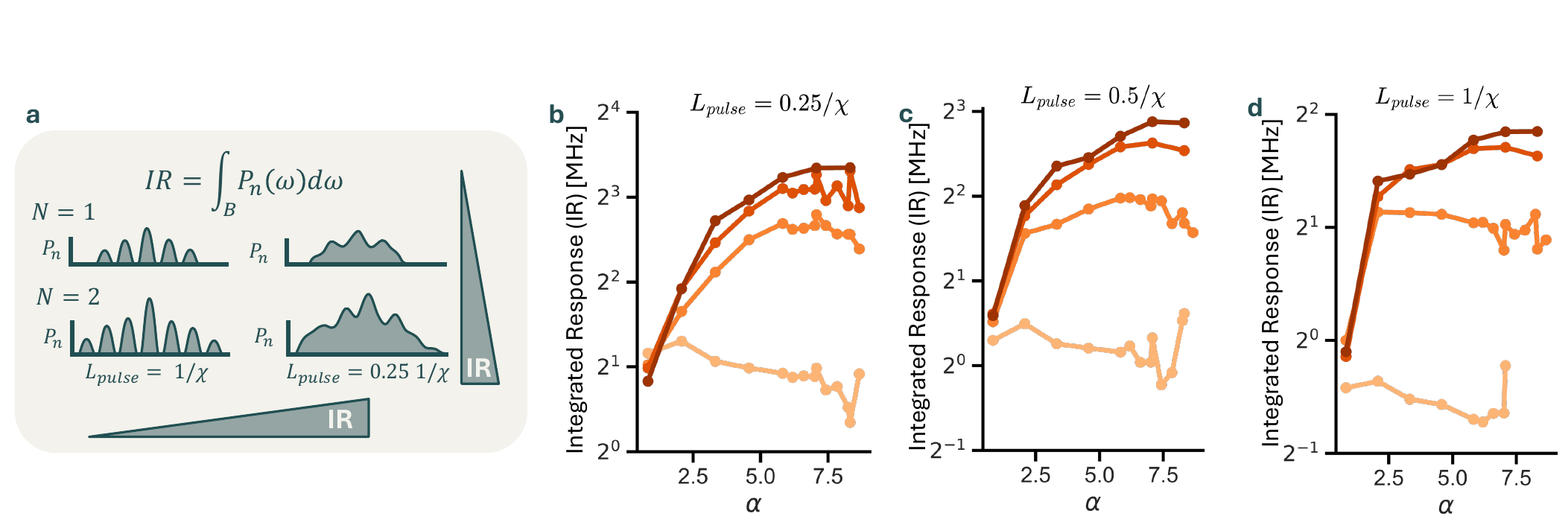}
    \caption{\textbf{Integrated sensing response for different pulse durations.}
    \textbf{(a)} Definition of the integrated response, $\mathrm{IR}=\int_B P_n(f)\,df$, which captures both the amplitude and spectral extent of the sensing response. 
    The schematic illustrates how the integrated response changes with Grover iteration number and sensing-pulse duration.
    \textbf{(b)--(d)} Integrated response as a function of the initial displacement amplitude $\alpha$ for $N=0$-$3$ Grover iterations and sensing-pulse durations $L_{\mathrm{pulse}}=0.25/\chi$, $0.5/\chi$, and $1/\chi$, respectively. 
    Grover amplification increases the integrated response over a broad range of $\alpha$, while longer sensing pulses reduce the total integrated response because of their increased spectral selectivity.
    }
    \label{fig:integrated signal}
\end{figure*}
Grover-based amplitude amplification increases the qubit-excitation probability for in-band signals while extending the effective detection bandwidth.
We quantify this combined enhancement by defining an integrated response (IR) that accounts for both the frequency range and the response strength.
We smooth the excitation-probability curve with a Gaussian filter of standard deviation $\sigma=2$ bins and estimate the local baseline $p_{\mathrm{bg}}$ as the mean over the two class-0 flanks.
We then compute

\begin{equation}
IR=\Delta f\sum_{j\in\mathcal{W}}\left[\tilde{p}_e(f_j)-p_{\mathrm{bg}}\right],
\end{equation}

where $\mathcal{W}$ is the detected window, $\tilde{p}_e$ is the smoothed excitation probability, and $\Delta f$ is the frequency spacing between bins. 
This quantity has units of frequency and accounts for both the width and contrast of the response. We exclude curves for which no valid window is detected, or whose window spans fewer than two bins or more than two-thirds of the measured curve.

Figure~\ref{fig:integrated signal}b shows that the $N=1$ and $N=2$ protocols achieve a larger IR than the coherent-probe baseline over a range of $\alpha$. We repeat the experiment for sensing-pulse durations of $1/(4\chi)$, $1/(2\chi)$, and $1/\chi$ and observe the same trend of increasing IR with the number of Grover iterations.
By amplifying initially weak Fock-state components, the Grover protocols allow a broader set of frequency channels to contribute to the qubit response.
This stronger and broader response improves the distinguishability of in-band and out-of-band signals, providing the physical basis for the observed classification advantage.
The gain in integrated response persists for longer, more spectrally selective sensing pulses.
For these pulse durations, the Grover protocols retain their resource advantage over the coherent-probe baseline, while the integrated response increases with the number of Grover iterations.
Longer sensing pulses yield better-resolved spectral peaks, reducing the IR, as shown in Fig.~\ref{fig:integrated signal}b--d.

\subsection{Ideal amplification response}
\label{app_sec:theory}

We consider ideal number-selective phase gates and neglect decoherence.
The cavity is initialized in a coherent state $\ket{\alpha}$ with mean photon number $\bar{n}=|\alpha|^2$. 
One Grover iteration is
\begin{equation}
    \begin{aligned}
    \hat{V}_i(\alpha)
    &=\hat{D}(\alpha)\hat{S}^{(0)}_\pi
      \hat{D}(-\alpha)\hat{S}^{(i)}_\pi \\
    &=\left(\hat{I}-2\ket{\alpha}\bra{\alpha}\right)
      \left(\hat{I}-2\ket{i}\bra{i}\right),
    \end{aligned}
\end{equation}
where $\hat{S}^{(i)}_\pi=\hat{I}-2\ket{i}\bra{i}$.
These two reflections amplify the targeted Fock-state population within the subspace spanned by $\ket{\alpha}$ and $\ket{i}$.
After $N$ iterations, this population is
\begin{equation}
    P_N(i,\alpha)
    =\sin^2\!\left[(2N+1)\theta_i\right],
    \qquad
    \theta_i=\arcsin\sqrt{p_i},
    \qquad
    p_i=\frac{e^{-\bar{n}}\bar{n}^{\,i}}{i!}.
    \label{eq:ideal_response}
\end{equation}
An ideal selective readout maps $P_N(i,\alpha)$ onto the qubit-excitation probability.

The iteration count determines both the amplification and the shape of the response.  
We define $N_{\mathrm{opt}}$ by the first amplification maximum of the most populated initial Fock state, $i_0=\lfloor\bar{n}\rfloor$. 
The condition $(2N+1)\theta_{i_0}=\pi/2$ gives
\begin{equation}
    N_{\mathrm{opt}}
    =\operatorname{round}\!\left(
    \frac{\pi}{4\theta_{i_0}}-\frac{1}{2}
    \right).
    \label{eq:optimal_iterations}
\end{equation}
Integer rounding generally prevents exact unit population.
Beyond this first maximum, the central response decreases while less populated neighboring states continue to amplify, producing a central dip.

\begin{figure*}
    \centering    \includegraphics[width=0.75\linewidth]{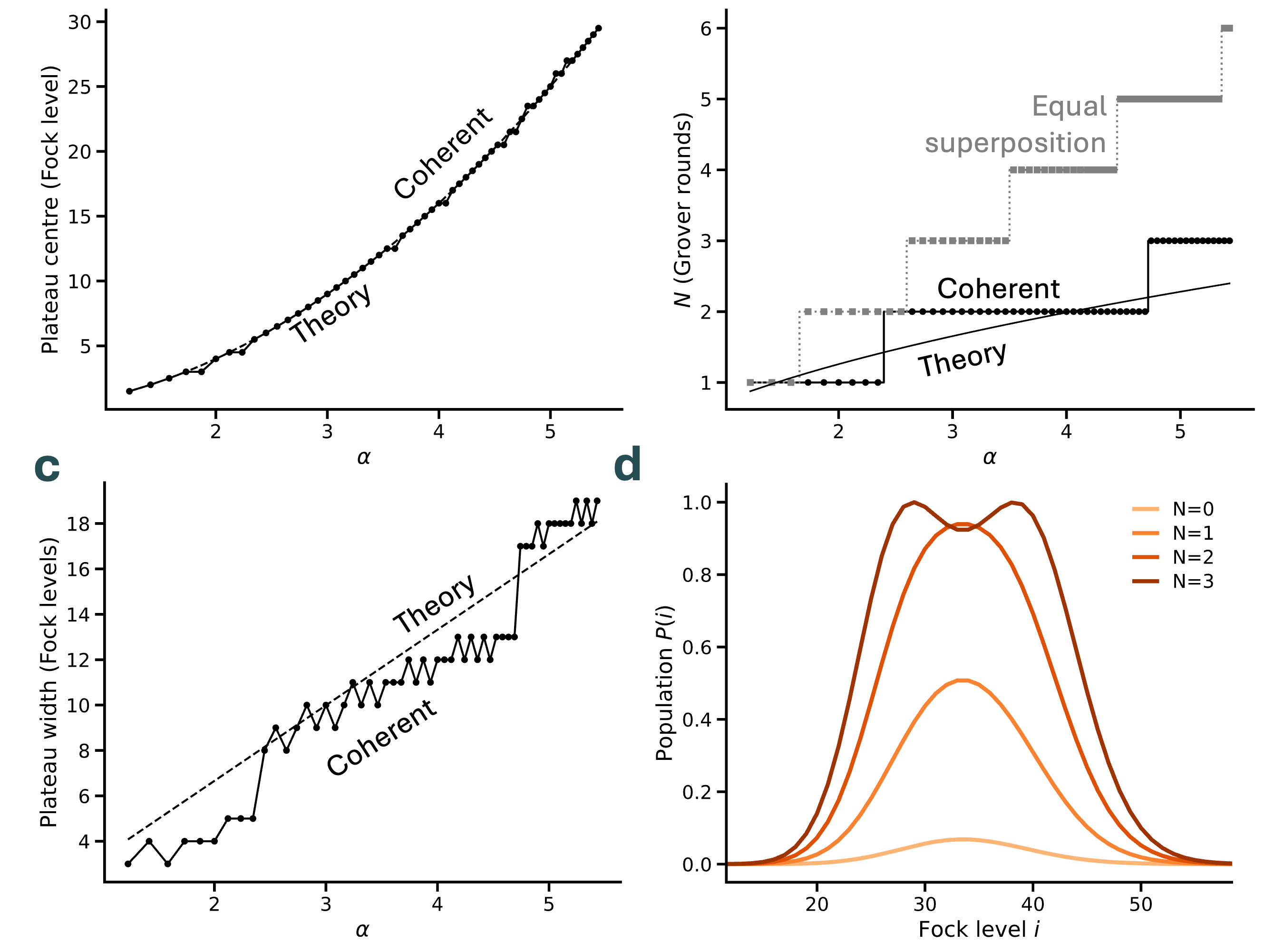}
    \caption{
    \textbf{Ideal response and iteration count for coherent-state Grover amplification.}
    (a) Center of the amplified response versus displacement amplitude
    $\alpha$, compared with $i_{\mathrm{c}}\simeq|\alpha|^2$.
    (b) Grover iteration count selected by the squareness criterion
    for a coherent initial state, compared with the theoretical
    first-maximum estimate and the count for an equal superposition
    with the same mean photon number.
    (c) Full width at half maximum of the amplified response,
    compared with $W_{\mathrm{FWHM}}\simeq4|\alpha|\sqrt{\ln2}$.
    (d) Response versus targeted Fock level $i$ for $N=0,1,2,3$,
    Additional iterations initially increase and flatten the response,
    while amplification beyond the first central maximum produces
    a central dip.
    }
\label{fig:supp_theory}
\end{figure*}

For $\bar{n}\gg1$, the Gaussian approximation to the Poisson distribution gives
\begin{equation}
    P_N(i,\alpha)\simeq
    \sin^2\!\left[
    A\,e^{-(i-\bar{n})^2/(4\bar{n})}
    \right],
    \qquad
    A=\frac{2N+1}{(2\pi\bar{n})^{1/4}}.
    \label{eq:response_envelope}
\end{equation}
The response is therefore approximately centered at $i_{\mathrm{c}}\simeq\bar{n}$, consistent with Fig.~\ref{fig:supp_theory}a.
The first central maximum corresponds to $A\simeq\pi/2$, yielding the continuous estimate
\begin{equation}
    N_{\mathrm{opt}}\simeq
    \frac{\pi}{4}(2\pi\bar{n})^{1/4}-\frac{1}{2}.
\end{equation}
Figure~\ref{fig:supp_theory}.b compares this estimate with the iteration count selected by the squareness criterion and the count for an equal superposition with the same mean photon number.
At this operating point, the response envelope has full width at half maximum
\begin{equation}
    W_{\mathrm{FWHM}}\simeq4\sqrt{\bar{n}\ln2},
\end{equation}
as compared with the calculated widths in Fig.~\ref{fig:supp_theory}c.
Thus, the displacement sets the response center through $i_{\mathrm{c}}\simeq|\alpha|^2$, while the width near optimal amplification grows as $|\alpha|$ and the required iteration count scales as $|\alpha|^{1/2}$.
Figure~\ref{fig:supp_theory}d illustrates the evolution of the response with $N$.

\end{document}